# A TAXONOMY OF SCHEDULERS – OPERATING SYSTEMS, CLUSTERS AND BIG DATA FRAMEWORKS

LESZEK SLIWKO (LSLIWKO@GMAIL.COM)
AXIS APPLICATIONS LTD, LONDON, UK

**Abstract**: This review analyzes deployed and actively used workload schedulers' solutions and presents a taxonomy in which those systems are divided into several hierarchical groups based on their architecture and design. While other taxonomies do exist, this review has focused on the key design factors that affect the throughput and scalability of a given solution, as well as the incremental improvements which bettered such an architecture. This review gives special attention to Google's Borg, which is one of the most advanced and published systems of this kind.

## 1. TAXONOMY OF SCHEDULERS

Although managing workload in a Cloud system is a modern challenge, scheduling strategies are a well-researched field as well as being an area where there has been considerable practical implementation. This background review started by analyzing deployed and actively used solutions and presents a taxonomy in which schedulers are divided into several hierarchical groups based on their architecture and design. While other taxonomies do exist (e.g., Krauter et al., 2002; Yu and Buyya, 2005; Pop et al., 2006; Smanchat and Viriyapant, 2015; Rodriguez and Buyya, 2017; Zakarya and Gillam, 2017; Tyagi and Gupta, 2018), this review has focused on the most important design factors that affect the throughput and scalability of a given solution, as well as the incremental improvements which bettered such an architecture.

Figure 1 visualizes how the schedulers' groups are split. The sections which follow discusses each of these groups separately.

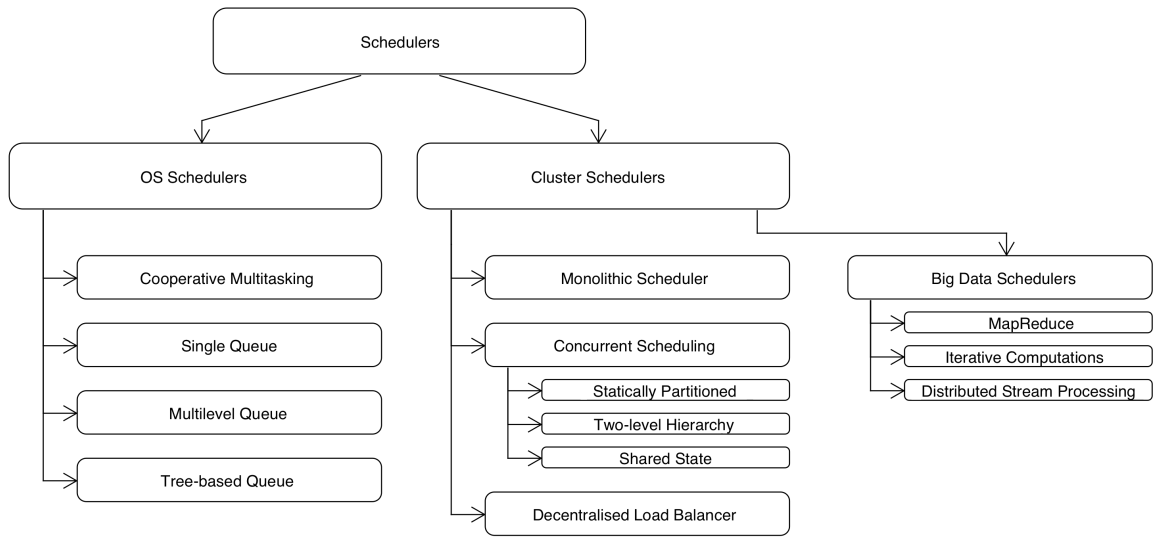

Figure 1: Schedulers taxonomy

## 2. METACOMPUTING

The concept of connecting computing resources has been an active area of research for some time. The term 'metacomputing' was established as early as 1987 (Smarr and Catlett, 2003) and since then the topic of scheduling has been the focus of many research projects, such as (i) service localizing idle workstations and utilizing their spare CPU cycles – HTCondor (Litzkow et al., 1988); (ii) the Mentat – a parallel run-time system developed at the University of Virginia (Grimshaw, 1990); (iii) blueprints for a national supercomputer (Grimshaw et al., 1994), and (iv) the Globus metacomputing infrastructure toolkit (Foster and Kesselman, 1997).

Before the work of Foster et al. (2001), there was no clear definition to what 'grid' systems referred. Following this publication, the principle that grid systems should allow a set of participants to share several connected computer machines and their resources became established. A list of rules defines these shared system policies. This includes which resources are being shared, who is sharing these resources, the extent to which they can use those resources, and what quality of service they can expect.

As shown in the following sections, the requirements of a load balancer in a decentralized system varies significantly compared to scheduling jobs on a single machine (Hamscher et al., 2000). One significant difference is the network resources, in that transferring data between machines is expensive because the nodes tend to be geographically distributed. In addition to the high-impact spreading of tasks across networked machines, the load balancer in Clusters generally provides a mechanism for fault-tolerance and user session management. The sections below also explain the workings of several selected current and historical schedulers and distributed frameworks. If we can understand these, we will know more about how scheduling algorithms developed over time, as well as the different ways they have been conceptualized. This paper does not purport to be a complete taxonomy of all available designs, but rather presents an analysis of some of the most important concepts and aspects of the history of schedulers.

## 3. OS SCHEDULERS

The Operating System (OS) Scheduler, also known as a 'short-term scheduler' or 'CPU scheduler', works within very short time frames, i.e., time-slices. During scheduling events, an algorithm must examine planned tasks and assign them appropriate CPU times (Bulpin, 2005; Arpaci-Dusseau and Arpaci-Dusseau, 2015). This setting requires schedulers to use highly optimized algorithms with very small overheads. Process schedulers face the challenge of how to maintain the balance between throughput and responsiveness (i.e., minimum latency). Prioritizing the execution of processes with a higher sleep/processing ratio is the way this is generally achieved (Pabla, 2009).

At present, the most advanced strategies also take into consideration the latest CPU core where the process ran the previous time, which is known as 'Non-Uniform Memory Access (NUMA) awareness'. The aim is to reuse the same CPU cache memory wherever possible (Blagodurov et al., 2010). The memory access latency differences can be very substantial, for example ca. 3-4 cycles for L1 cache, ca. 6-10 cycles for L2 cache and ca. 40-100 cycles for L3 cache (Drepper, 2007). NUMA awareness also involves prioritizing the act of choosing a real idle core which must occur before its logical SMT sibling, also known as 'Hyper-Threading (HT) awareness'. Given this, NUMA awareness is a crucial element in the design of modern OS schedulers. With a relatively high data load to examine in a short period, implementation needs to be strongly optimized to ensure faster execution.

OS Schedulers tend to provide only a very limited set of configurable parameters, wherein the access to modify them is not straightforward. Some of the parameters can change only during the kernel compilation process and require rebooting, such as compile-time options CONFIG_FAIR_USER_SCHED and CONFIG_FAIR_CGROUP_SCHED, or on the fly using the low-level Linux kernel's tool 'sysctl'.

### 3.1. COOPERATIVE MULTITASKING

Early multitasking Operating Systems, such as Windows 3.1x, Windows 95, 96 and Me, Mac OS before X, adopted a concept known as Cooperative Multitasking or Cooperative Scheduling (CS). In early implementations of CS, applications voluntarily ceded CPU time to one another. This was later supported natively by the OS, although Windows 3.1x used a non-pre-emptive scheduler which did not interrupt the program, wherein the program needed to explicitly tell the system that it no longer required the processor time. Windows 95 introduced a rudimentary pre-emptive scheduler, although this was for 32-bit applications only (Hart, 1997). The main issue in CS is the hazard caused by the poorly designed program. CS relies on processes regularly giving up control to other processes in the system, meaning that if one process consumes all the available CPU power then all the systems will hang.

### 3.2. SINGLE QUEUE

Before Linux kernel version 2.4, the simple Circular Queue (CQ) algorithm was used to support the execution of multiple processes on the available CPUs. A Round Robin policy informed the next process run (Shreedhar, 1995). In kernel version 2.2, processes were further split into non-real/real-time categories, and scheduling classes were introduced. This algorithm was replaced by O(n) scheduler in Linux kernel versions 2.4-2.6. In O(n), processor time is divided into epochs, and within each epoch every task can execute up to its allocated time slice before being pre-empted. At the beginning of each epoch, the time slice is given to each task; it is based on the task's static priority added to half of any remaining time-slices from the previous epoch (Bulpin, 2005). Thus, if a task does not use its entire time slice in the current epoch, it can execute for longer in the next one. During a scheduling event, an O(n) scheduler requires iteration through all the process which are currently planned (Jones, 2009), which can be seen as a weakness, especially for multi-core processors.

Between Linux kernel versions 2.6-2.6.23 came the implementation of the O(1) scheduler. O(1) further splits the processes list into active and expired arrays. Here, the arrays are switched once all the processes from the active array have exhausted their allocated time and have been moved to the expired array. The O(1) algorithm analyses the average sleep time of the process, with more interactive tasks being given higher priority to boost system responsiveness. The calculations required are complex and subject to potential errors, where O(1) may cause non-interactive behavior from an interactive process (Wong et al., 2008; Pabla, 2009).

### 3.3. MULTILEVEL QUEUE

With Q(n) and O(1) algorithms failing to efficiently support the applications' interactivity, the design of OS Scheduler evolved into a multilevel queue. In this queue, repeatedly sleeping (interactive) processes are pushed to the top and executed more frequently. Simultaneously, background processes are pushed down and run less frequently, although for extended periods.

Perhaps the most widespread scheduler algorithm is Multilevel Feedback Queue (MLFQ), which is implemented in all modern versions of Windows NT (2000, XP, Vista, 7 and Server), Mac OS X, NetBSD and Solaris kernels (up to version 2.6, when it was replaced with O(n) scheduler). MLFQ was first described in 1962 in a system known as the Compatible Time-Sharing System (Corbató et al., 1962). Fernando Corbató was awarded the Turing Award by the ACM in 1990 'for his pioneering work organizing the concepts and leading the development of the general-purpose, large-scale, time-sharing and resource-sharing computer systems, CTSS and Multics'. MLFQ organizes jobs into a set of queues $Q_0, Q_1, ..., Q_i$ wherein a job is promoted to a higher queue if it does not finish within $2^i$ time units. The algorithm always processes the job from the front of the lowest queue, meaning that short processes have preference. Although it has a very poor worst-case scenario, MLFQ turns out to be very efficient in practice (Becchetti et al., 2006).

Staircase Scheduler (Corbet, 2004), Staircase Deadline Scheduler (Corbet, 2007), Brain F. Scheduler (Groves et al., 2009) and Multiple Queue Skiplist Scheduler (Kolivas, 2016) constitute a line of successive schedulers developed by Con Kolivas since 2004 which are based on a design of Fair Share Scheduler from Kay and Lauder (1988). None of these schedulers have been merged into the source code of mainstream kernels. They are available only as experimental '-ck' patches. Although the concept behind those schedulers is similar to MLFQ, the implementation details differ significantly. The central element is a single, ranked array of processes for each CPU ('staircase'). Initially, each process (P1, P2, ...) is inserted at the rank determined by its base priority; the scheduler then picks up the highest ranked process (P) and runs it. When P has used up its time slice, it is reinserted into the array but at a lower rank, where it will continue to run but at a lower priority. When P exhausts its next time-slice, its rank is lowered again. P then continues until it reaches the bottom of the staircase, at which point it is moved up to one rank below its previous maximum and is assigned two time-slices. When P exhausts these two time-slices, it is reinserted once again in the staircase at a lower rank. When P once again reaches the bottom of the staircase, it is assigned another time-slice and the cycle repeats. P is also pushed back up the staircase if it sleeps for a predefined period. The result of this is that that interactive tasks which tend to sleep more often should remain at the top of the staircase, while CPU-intensive processes should continuously expend more time-slices but at a lower frequency. Additionally, each rank level in the staircase has its quota, and once the quota is expired all processes on that rank are pushed down.

Most importantly, Kolivas' work introduced the concept of 'fairness'. What this means is that each process gets a comparable share of CPU time to run, proportional to the priority. If the process spends much of its time waiting for I/O events, then its spent CPU time value is low, meaning that it is automatically prioritized for execution. When this happens, interactive tasks which spend most of their time waiting for user input get execution time when they need it, which is how the term 'sleeper fairness' derives. This design also prevents a situation in which the process is 'starved', i.e., never executed.

### 3.4. TREE-BASED QUEUE

While the work of Con Kolivas has never been merged into the mainstream Linux kernel, it has introduced the central concept of 'fairness', which is the crucial feature of the design of most current OS schedulers. At the time of writing, Linux kernel implements Completely Fair Scheduler (CFS), which was developed by Ingo Molnár and introduced in kernel version 2.6.23. A central element in this algorithm is a self-balancing red-black tree structure in which processes are indexed by spent processor time. CFS implements the Weighted Fair Queueing (WFQ) algorithm, in which the available CPU time-slices are split between processes in proportion to their priority weights ('niceness'). WFQ is based on the idea of the 'ideal processor', which means that each process should have an equal share of CPU time adjusted for their priority and total CPU load (Jones, 2009; Pabla, 2009).

Lozi et al. (2016) presents an in-depth explanation of the algorithm's workings, noting potential issues regarding the CFS approach. The main criticism revolves around the four problematic areas:

- Group Imbalance – the authors' experiments have shown that not every core of their 64-core machine is equally loaded: some cores run only one process or sometimes no processes at all, while the rest of the cores were overloaded. It seems that the scheduler was not balancing the load because of the hierarchical design and complexity of the load tracking metric. To limit the complexity of the scheduling algorithm, the CPU cores are grouped into scheduling groups, i.e., nodes. When an idle core attempts to steal work from another node, it compares only the average load of its node with that of its victim's node. It will steal work only if the average load of its victim's group is higher than its own. The result is inefficiency since idle cores will be concealed by their nodes' average load.
- Scheduling Group Construction – this concern relates to the way scheduling groups are constructed which is not adapted to modern NUMA machines. Applications in Linux can be pinned to a subset of available cores. CFS might assign the same cores to multiple scheduling groups with those groups then being ranked by distance. This could be nodes one hop apart, two hops apart and so on. This feature was designed to increase the probability that processes would remain close to their original NUMA node. However, this could result in the application being pinned to particular cores which are separated by more than one hop, with work never being migrated outside the initial core. This might mean that an application uses only one core.
- Overload-on-Wakeup – this problem occurs when a process goes to sleep on a particular node and is then awoken by a process on the same node. In such a scenario, only cores in this scheduling group will be considered to run this process. The aim of this optimization is to improve cache utilization by running a process close to the waker process, meaning that there is the possibility of them sharing the last-level memory cache. However, the might be the scheduling of a process on a busy core when there are idle cores in alternative nodes, resulting in the severe underutilization of the machine.
- Missing Scheduling Domains – this is the result of a line of code omission while refactoring the Linux kernel source code. The number of scheduling domains is incorrectly updated when a

> particular code is disabled and then enabled, and a loop exits early. As a result, processes can be run only on the same scheduling group as their parent process.

Lozi et al. (2016) have provided a set of patches for the above issues and have presented experimental results after applying fixes. They have also made available a set of tools on their site which could be used to detect those glitches early in the Linux kernel lifecycle. Moreover, it has been argued that the sheer number of optimizations and modifications implemented into CFS scheduler changed the initially simple scheduling policy into one which was very complex and bug-prone. As of 12$^{th}$ February 2019, there had been 780 commits to CFS source code ('fair.c' file in github.com/torvalds/linux repository) since November 2011. As such, an alternative approach is perhaps required, such as a scheduler architecture based on pluggable components. This work demonstrates the immerse complexity of scheduling solutions catering to the complexities of modern hardware.

## 4. CLUSTER SCHEDULERS

There are many differences between distributed computing and traditional computing. For example, the physical size of the system means that there may be thousands of machines involved, with thousands of users being served and millions of API calls or other requests needing processing. While responsiveness and low overheads are often the focus of process schedulers, the focus of cluster schedulers is to focus upon high throughput, fault-tolerance, and scalability. Cluster schedulers usually work with queues of jobs spanning to hundreds of thousands, and indeed sometimes even millions of jobs. They also seem to be more customized and tailored to the needs of the organization which is using them.

Cluster schedulers often provide complex administration tools with a wide spectrum of configurable parameters and flexible workload policies. All configurable parameters can generally be accessed via configuration files or the GUI interface. However, it appears that site administrators seldom stray from a default configuration (Etsion and Tsafrir, 2005). The most used scheduling algorithm is simply a First-Come-First-Serve (FCFS) strategy with backfilling optimization.

The most common issues which cluster schedulers must deal with are:

- Unpredictable and varying load (Moreno et al., 2013);
- Mixed batch jobs and services (ibid.);
- Complex policies and constraints (Adaptive Computing, 2002);
- Fairness (ibid.);
- A rapidly increasing workload and cluster size (Isard et al., 2007);
- Legacy software (ibid.);
- Heterogeneous nodes with a varying level of resources and availability (Thain et al., 2005);
- The detection of underperforming nodes (Zhang et al., 2014);
- Issues related to fault-tolerance (ibid.) and hardware malfunctions (Gabriel et al., 2004).

Another challenge, although one which is rarely tackled by commercial schedulers, is minimizing total power consumption. Typically, idle machines consume around half of their peak power (McCullough et al., 2011). Therefore, a Data Center can decrease the total power it consumes by concentrating tasks on fewer machines and powering down the remaining nodes (Pinheiro et al., 2001; Lang and Patel, 2010).

The proposed grouping of Cluster schedulers loosely follows the taxonomy presented in Schwarzkopf et al. (2013).

## 4.1. MONOLITHIC SCHEDULER

The earliest Cluster schedulers had a centralized architecture in which a single scheduling policy allocated all incoming jobs. The tasks would be picked from the head of the queue and scheduled on system nodes in a serial manner by an allocation loop. Examples of centralized schedulers include Maui (Jackson et al., 2001) and its successor Moab (Adaptive Computing, 2015), Univa Grid Engine (Gentzsch, 2001), Load Leveler (Kannan et al., 2001), Load Sharing Facility (Etsion and Tsafrir, 2005), Portable Batch System (Bode et al., 2000) and its successor TORQUE (Klusáček et al., 2013), Alibaba's Fuxi (Zhang et al., 2014), Docker Swarm (Naik, 2016), Kubernetes (Vohra, 2017) and several others.

Monolithic schedulers implement a wide array of policies and algorithms, such as FCFS, FCFS with backfilling and gang scheduling, Shortest Job First (SJF), and several others. The Kubernetes (Greek: 'κυβερνήτης') scheduler implements a range of scoring functions such as node or pod affinity/anti-affinity, resources best-fit and worst-fit, required images locality, etc. which can be additionally weighted and combined into node's score values (Lewis and Oppenheimer, 2017). As an interesting note – one of the functions (BalancedResourceAllocation routine) implemented in Kubernetes evaluates the balance of utilized resources (CPU and memory) on a scored node.

Monolithic schedulers often face a 'head-of-queue' blocking problem, in which shorter jobs are held when a long job is waiting for a free node. To try and counter this problem, the schedulers often implement 'backfilling' optimization, where shorter jobs are allowed to execute while the long job is waiting. Perhaps the most widespread scheduler is Simple Linux Utility for Resource Management (SLURM) (Yoo et al., 2003). SLURM uses a best-fit algorithm which is based on either Hilbert curve scheduling or fat tree network topology; it can scale to thousands of CPU cores (Pascual, 2009). At the time of writing, the fastest supercomputer in the world is Sunway TaihuLight (Chinese: '神威·太湖之光'), which uses over 40k CPU processors, each of which contains 256 cores. Sunway TaihuLight's workload in managed by SLURM (TOP500 Project, 2017).

The Fuxi (Chinese: '伏羲') scheduler presents a unique strategy in that it matches newly-available resources against the backlog of tasks rather than matching tasks to available resources on nodes. This technique allowed Fuxi to achieve very high utilization of Cluster resources, namely 95% utilization of memory and 91% utilization of CPU. Fuxi has been supporting Alibaba's workload since 2009, and it scales to ca. 5k nodes (Zhang et al., 2014).

While Cluster scheduler designs have generally moved towards solutions which are more parallel, as demonstrated in the next subsection, centralized architecture is still the most common approach in High-Performance Computing. Approximately half the world's supercomputers use SLURM as their workload manager, while Moab is currently deployed on about 40% of the top 10, top 25 and top 100 on the TOP500 list (TOP500 Project, 2017).

## 4.2. CONCURRENT SCHEDULING

Historically, monolithic schedulers were frequently built on the premise of supporting a single 'killer-application' (Barroso et al., 2003). However, the workload of the data center has become more heterogeneous as systems and a modern Cluster system runs hundreds of unique programs with distinctive resource requirements and constraints. A single code base of centralized workload manager means that it is not easy to add a variety of specialized scheduling policies. Furthermore, as workload size is increased, the time to reach a scheduling decision is progressively limited. The result of this is a restriction in the selection of scheduling algorithms to less sophisticated ones, which affects the quality of allocations. To tackle those challenges, the Cluster schedulers developed designs which are more parallel.

### 4.2.1. STATICALLY PARTITIONED

The solution to the numerous policies and the lack of parallelism in central schedulers was to split Cluster into specialized partitions and manage them separately. Quincy (Isard et al., 2009), a scheduler managing workload of Microsoft's Dryad, follows this approach.

The development of an application for Dryad is modeled as a Directed Acyclic Graph (DAG) model in which the developer defines an application dataflow model and supplies subroutines to be executed at specified graph vertices. The scheduling policies and tuning parameters are specified by adjusting weights and capacities on a graph data structure. The Quincy implements a Greedy strategy. In this approach, the scheduler assumes that the currently scheduled job is the only job running on a cluster and so always selects the best node available. Tasks are run by remote daemon services. From time to time these services update the job manager about the execution status of the vertex, which in the case of failure might be re-executed. Should any task fail more than a configured number of times, the entire job is marked as failed (Isard et al., 2007).

Microsoft has built several frameworks on top of Dryad, such as COSMOS (Helland and Harris, 2011) which provided SQL-like language optimized for parallel execution. COSMOS was designed to support data-driven search and advertising within the Windows Live services owned by Microsoft, such as Bing, MSN, and Hotmail. It analyzed user behaviors in multiple contexts, such as what people searched for, what links they clicked, what sites they visited, the browsing order, and the ads they clicked on. Although the Dryad project had several preview releases, it was eventually dropped when Microsoft shifted its focus to the development of Hadoop.

The main criticism of the static partitioning is inflexibility, that is, the exclusive sets of machines in a Cluster are dedicated to certain types of workload. That might result in a part of scheduler being relatively idle, while other nodes are very active. This issue leads to the Cluster's fragmentation and the suboptimal utilization of available nodes since no machine sharing is allowed.

### 4.2.2. TWO-LEVEL HIERARCHY

The solution to the inflexibility of static partitioning was to introduce two-level architecture in which a Cluster is partitioned dynamically by a central coordinator. The actual task allocations take place at the second level of architecture in one of the specialized schedulers. The first two-level scheduler was Mesos (Hindman et al., 2011). It was developed at the University of California (Berkeley) and is now hosted in the Apache Software Foundation. Mesos was a foundation base for other Cluster systems such as Twitter's Aurora (Aurora, 2018) and Marathon (Mesosphere, 2018).

Mesos introduces a two-level scheduling mechanism in which a centralized Mesos Master acts as a resource manager. It dynamically allocates resources to different scheduler frameworks via Mesos Agents, e.g., Hadoop, Spark and Kafka. Mesos Agents are deployed on cluster nodes and use Linux's cgroups or Docker container (depending upon the environment) for resource isolation. Resources are distributed to the frameworks in the form of 'offers' which contain currently unused resources. Scheduling frameworks have autonomy in deciding which resources to accept and which tasks to run on them.

Mesos is most effective when tasks are relatively small, short-lived and have a high resource churn rate, i.e., they relinquish resources more frequently. In the current version (1.4.1), only one scheduling framework can examine a resource offer at any given time. This resource is effectively locked for the duration of a scheduling decision, meaning that concurrency control is pessimistic. Campbell (2017) presents several practical considerations for using Mesos in the production environment, in addition to advice on best practice.

Two-level schedulers offered a working solution to the lack of parallelization found in central schedulers and the low efficiency of statically partitioned Clusters. Nevertheless, the mechanism used causes resources to remain locked at the same time a specialized scheduled examines the resources offer. This means the benefits from parallelization are limited due to pessimistic locking. Furthermore, the schedulers do not coordinate with each other and must rely on a centralized coordinator to make them offers. This further restricts their visibility of the resources in a Cluster.

### 4.2.3. SHARED STATE

To address the limited parallelism of the two-level scheduling design, the alternative approach taken by some organizations was to redesign schedulers' architecture into several schedulers, all working concurrently. The schedulers work on a shared Cluster's state information and manage their resources' reservations using an optimistic concurrency control method. A sample of such systems includes: Microsoft's Apollo (Boutin et al., 2014); Omega, Google Borg's spinoff (Schwarzkopf et al., 2013); HashiCorp's Nomad (HashiCorp, 2018); and also Borg (Burns et al., 2016) itself. The latter system has been refactored from monolithic into parallel architecture after experimentations with Omega.

By default, Nomad runs one scheduling worker per CPU core. Scheduling workers pick job submissions from the broker queue and then submit it to one of the three schedulers: a long-lived services scheduler, a short-lived batch jobs scheduler and a system scheduler, which is used to run internal maintenance routines. Additionally, Nomad can be extended to support custom schedulers. Schedulers process and generate an action plan, which constitutes a set of operations to create new allocations, or to evict and update existing ones (HashiCorp, 2018).

Microsoft's Apollo design seems to be primarily tuned for high tasks churn, and at peak times is capable of handling more than 100k of scheduling requests per second on a ca. 20k nodes cluster. Apollo uses a set of per-job schedulers called Job Managers (JM) wherein a single job entity contains a multiplicity of tasks which are then scheduled and executed on computing nodes. Tasks are generally short-lived batch jobs (Boutin et al., 2014). Apollo has a centralized Resource Monitor (RM), while each node runs its Process Node (PN) with a queue of tasks. Each PN is responsible for local scheduling decisions and can independently reorder its job queue to allow smaller tasks to be executed immediately, while larger tasks wait for resources to become available. In addition, PN computes a wait-time matrix based on its queue which publicizes the future availability of the node's resources. Scheduling decisions are made optimistically by JMs based on the shared cluster's resource state, which is continuously retrieved and aggregated by RM.

Furthermore, Apollo categorizes tasks as 'regular' and 'opportunistic'. Opportunistic tasks are used to fill resource gaps left by regular tasks. The system also prevents overloading the cluster by limiting the total number of regular tasks that can be run on a cluster. Apollo implements locality optimization by taking into consideration the location of data for a given task. For example, the system will score nodes higher if the required files are already on the local drive as opposed to machines needing to download data (Boutin et al., 2014).

Historically, Omega was a spinoff from Google's Borg scheduler. Despite the various optimizations acquired by Borg over the years, including internal parallelism and multi-threading, to address the issues of head-of-line blocking and scalability problems, Google decided to create an Omega scheduler from the ground up (Schwarzkopf et al., 2013). Omega introduced several innovations, such as storing the state of the cluster in a centralized Paxos-based store that was accessed by multiple components simultaneously. Optimistic locking concurrency control resolved the conflicts which emerged. This feature allowed Omega to run several schedulers at the same time and improve the throughput. Many of Omega's innovations have since been folded into Borg (Burns et al., 2016).

Omega's authors highlight the disadvantages of the shared state and parallel reservation of resources, namely: (i) the state of a node could have changed considerably when the allocation decision was being made, and it is no longer possible for this node to accept a job; (ii) two or more allocations to the same node could have conflicted and both scheduling decisions are nullified; and (iii) this strategy introduces significant difficulties when gang-scheduling a batch of jobs as (i) or (ii) are happening (Schwarzkopf et al., 2013).

In this research, Google's Borg received special attention, as one of the most advanced and published schedulers. Moreover, while other schedulers are designed to support either a high churn of short-term jobs, e.g., Microsoft's Apollo (Boutin et al., 2014), Alibaba's Fuxi (Zhang et al., 2014), or else a limited number of long-term services, such as Twitter's Aurora (Aurora, 2018), Google's engineers have created a system which supports a mixed workload. Borg has replaced two previous systems, Babysitter and the Global Work Queue, which were used to manage long-running services and batch jobs separately (Burns et al., 2016). Given the significance of Borg's design for this research, it is discussed separately in section 5.

### 4.2.4. DECENTRALISED LOAD BALANCER

The research (Sliwko, 2018) proposes a new type of Cluster's workload orchestration model in which the actual scheduling logic is processed on nodes themselves. This is a significant step towards completely decentralized Cluster orchestration. The cluster state is retrieved from a subnetwork of BAs, although this system does not rely on the accuracy of this information and uses it exclusively to retrieve an initial set of candidate nodes where a task could potentially run. The actual task to machine matching is performed between the nodes themselves. As such, this design avoids the pitfalls of the concurrent resource locking, which includes conflicting scheduling decisions and the non-current state of nodes' information. Moreover, the decentralization of the scheduling logic also lifts complexity restrictions on scheduling logic, meaning that a wider range of scheduling algorithms can be used, such as metaheuristic methods.

### 4.3. BIG DATA SCHEDULERS

In taxonomy presented in this paper, Big Data schedulers are visualized as a separate branch from Cluster Schedulers. Although Big Data Schedulers seem to belong to one of the Cluster schedulers designs discussed previously, this separation signifies their over-specialization, and that only a very restricted set of operations is supported (Isard et al., 2007; Zaharia et al., 2010). The scheduling mechanisms are often intertwined with the programming language features, with Big Data frameworks often providing their own API (Zaharia et al., 2009; White, 2012) and indeed sometimes even their own custom programming language, as seen with Skywriting in CIEL (Murray et al., 2011).

Generally speaking, Big Data frameworks provide very fine-grained control over how data is accessed and processed over the cluster, such as Spark RDD objects persist operations or partitioners (Zaharia et al., 2012). Such a deep integration of scheduling logic with applications is a distinctive feature of Big Data technology. At the time of writing, Big Data is also the most active distributed computing research area, with new technologies, frameworks and algorithms being released regularly.

Big Data is the term which describes the storage and processing of any data sets so large and complex that they become unrealistic to process using traditional data processing applications based on relational database management systems. It depends on the individual organization as to how much data is described as Big Data. The following examples provide an idea of scale:

- The NYSE (The New York Stock Exchange) produces about 15 TB of new trade data per day (Singh, 2017);
- Facebook warehouse stores upwards of 300 PB of data, with an incoming daily rate of about 600 TB (Vagata and Wilfong, 2014);
- The Large Hadron Collider (Geneva, Switzerland) produces about fifteen petabytes of data per year (White, 2012).

As a result of a massive size of the stored and processed data, the central element of a Big Data framework is its distributed file system, such as Hadoop Distributed File System (Gog, 2012), Google File System (Ghemawat et al., 2003) and its successor Colossus (Corbett et al., 2013). The nodes in a Big Data cluster fulfill the dual purposes of storing the distributed file system parts, usually in a few replicas for fault-tolerance means, and also providing a parallel execution environment for system tasks. The speed difference between locally-accessed and remotely stored input data is very substantial, meaning that Big Data schedulers are very focused on providing 'data locality', which means running a given task on a node where input data are stored or are in the closest proximity to it.

The origins of the Big Data technology are in the 'MapReduce' programming model, which implements the concept of Google's inverted search index. Developed in 2003 (Dean and Ghemawat, 2010) and later patented in 2010 (U.S. Patent 7,650,331), the Big Data design has evolved significantly in the years since. It is presented in the subsections below.

### 4.3.1. MAPREDUCE

MapReduce is the most widespread principle which has been adopted for processing large sets of data in parallel. Originally, the name MapReduce only referred to Google's proprietary technology, but the term is now broadly used to describe a wide range of software, such as Hadoop, CouchDB, Infinispan, and MongoDB. The most important features of MapReduce are its scalability and fine-grained fault-tolerance. The 'map' and 'reduce' operations present in Lisp and other functional programming languages inspired the original thinking behind MapReduce (Dean and Ghemawat, 2010):

- 'Map' is an operation used in the first step of computation and is applied to all available data that performs the filtering and transforming of all key-value pairs from the input data set. The 'map' operation is executed in parallel on multiple machines on a distributed file system. Each 'map' task can be restarted individually, and a failure in the middle of a multi-hour execution does not require restarting the whole job from scratch.
- The 'Reduce' operation is executed after the 'map' operations complete. It performs finalizing operations, such as counting the number of rows matching specified conditions and yielding fields frequencies. The 'Reduce' operation is fed using a stream iterator, thereby allowing the framework to process the list of items one at the time, thus ensuring that the machine memory is not overloaded (Dean and Ghemawat, 2010; Gog, 2012).

Following the development of the MapReduce concept, Yahoo! engineers began the Open Source project Hadoop. In February 2008, Yahoo! announced that its production search index was being generated by a 10k-core Hadoop cluster (White, 2012). Subsequently, many other major Internet companies, including Facebook, LinkedIn, Amazon and Last.fm, joined the project and deployed it within their architectures. Hadoop is currently hosted in the Apache Software Foundation as an Open Source project.

As in Google's original MapReduce, Hadoop's users submit jobs which consist of 'map' and 'reduce' operation implementations. Hadoop splits each job into multiple 'map' and 'reduce' tasks. These tasks subsequently process each block of input data, typically 64MB or 128MB (Gog, 2012). Hadoop's scheduler allocates a 'map' task to the closest possible node to the input data required – so-called 'data locality' optimization. In so doing, we can see the following allocation order: the same node, the same rack and finally a remote rack (Zaharia et al., 2009). To further improve performance, the Hadoop framework uses 'backup tasks' in which a speculative copy of a task is run on a separate machine. The purpose of this is to finish the computation more quickly. If the first node is available but behaving poorly, it is known as a 'straggler', with the result that the job is as slow as the misbehaving task. This behavior can occur for many reasons, such as faulty hardware or misconfiguration. Google estimated that using 'backup tasks' could improve job response times by 44% (Dean and Ghemawat, 2010).

At the time of writing, Hadoop comes with a selection of schedulers, as outlined below:

- 'FIFO Scheduler' is a default scheduling system in which the user jobs are scheduled using a queue with five priority levels. Typically, jobs use the whole cluster, so they must wait their turn. When another job scheduler chooses the next job to run, it selects jobs with the highest priority, resulting in low-priority jobs being endlessly delayed (Zaharia et al., 2009; White, 2012).
- 'Fair Scheduler' is part of the cluster management technology Yet Another Resource Negotiator (YARN) (Vavilapalli et al., 2013), which replaced the original Hadoop engine in 2012. In Fair Scheduler, each user has their own pool of jobs, and the system focuses on giving each user a proportional share of cluster resources over time. The scheduler uses a version of 'max-min fairness' (Bonald et al., 2006) with minimum capacity guarantees that are specified as the number of 'map' and 'reduce' task slots to allocate tasks across users' job pools. When one pool is idle, and the minimum share of the tasks slots is not being used, other pools can use its available task slots.
- 'Capacity Scheduler' is the second scheduler introduced within the YARN framework. Essentially, this scheduled is a number of separate MapReduce engines, which contains FCFS scheduling for each user or organization. Those queues can be hierarchical, with a queue having children queues, and with each queue being allocated task slots capacity which can be divided into 'map' and 'reduce' tasks. Task slots allocation between queues is similar to the sharing mechanism between pools found in Fair Scheduler (White, 2012).

The main criticism of MapReduce is the acyclic dataflow programming model. The stateless 'map' task must be followed by a stateless 'reduce' task, which is then executed by the MapReduce engine. This model makes it challenging to repeatedly access the same dataset, a common action during the execution of iterative algorithms (Zaharia et al., 2009).

### 4.3.2. ITERATIVE COMPUTATIONS

Following the success of Apache Hadoop, several alternative designs were created to address Hadoop's suboptimal performance when running iterative MapReduce jobs. Examples of such systems include HaLoop (Bu et al., 2010) and Spark (Zaharia et al., 2010).

HaLoop has been developed on top of Hadoop, with various caching mechanisms and optimizations added. This makes the framework loop-aware, for example by adding programming support for iterative application and storing the output data on the local disk. Additionally, HaLoop's scheduler keeps a record of every data block processed by each task on physical machines. It considers inter-iteration locality when scheduling tasks which follow. This feature helps to minimize costly remote data retrieval, meaning that tasks can use data cached on a local machine (Bu et al., 2010; Gog, 2012).

Similar to HaLoop, Spark's authors noted a suboptimal performance of iterative MapReduce jobs in the Hadoop framework. In certain kinds of application, such as iterative Machine Learning algorithms and interactive data analysis tools, the same data are repeatedly accessed in multiple steps and then discarded; therefore, it does not make sense to send it back and forward to a central node. In such scenarios, Spark will outperform Hadoop (Zaharia et al., 2012).

Spark is built on top of HDSF, but it does not follow the two-stage model of Hadoop. Instead, it introduces resilient distributed datasets (RDD) and parallel operations on these datasets (Gog, 2012):

- 'reduce' - combines dataset elements using a provided function;
- 'collect' - sends all the elements of the dataset to the user program;
- 'foreach' - applies a provided function onto every element of a dataset.

Spark provides two types of shared variables:

- 'accumulators' - variables onto each worker can apply associative operations, meaning that they are efficiently supported in parallel;
- 'broadcast variables' - sent once to every node, with nodes then keeping a read-only copy of those variables (Zecevic, 2016).

The Spark job scheduler implementation is conceptually similar to that of Dryad's Quincy. However, it considers which partitions of RDD are available in the memory. The framework then re-computes missing partitions, and tasks are sent to the closest possible node to the input data required (Zaharia et al., 2012).

Another significant feature implemented in Spark is the concept of 'delayed scheduling'. In situations when a head-of-line job that should be scheduled next cannot launch a local task, Spark's scheduler delays the task execution and lets other jobs start their tasks instead. However, if the job has been skipped long enough, typically a period of up to ten seconds, it launches a non-local task. Since a typical Spark workload consists of short tasks, meaning that it has a high task slots churn, tasks have a higher chance of being executed locally. This feature helps to achieve 'data locality' which is nearly optimal, and which has a very small effect on fairness; in addition, the cluster throughput can be almost doubled, as shown in an analysis performed on Facebook's workload traces (Zaharia et al., 2010).

### 4.3.3. DISTRIBUTED STREAM PROCESSING

The core concept behind distributed stream processing engines is the processing of incoming data items in real time by modelling a data flow in which there are several stages which can be processed in parallel. Other techniques include splitting the data stream into multiple sub-streams and redirecting them into a set of networked nodes (Liu and Buyya, 2017).

Inspired by Microsoft's research into DAG models (Isard et al., 2009), Apache Storm (Storm) is a distributed stream processing engine used by Twitter following extensive development (Toshniwal et al., 2014). Its initial release was 17 September 2011, and by September 2014 it had become open-source and an Apache Top-Level Project.

The defined topology acts as a distributed data transformation pipeline. The programs in Storm are designed as a topology in the shape of DAG, consisting of 'spouts' and 'bolts':

- 'Spouts' read the data from external sources and emit them into the topology as a stream of 'tuples'. This structure is accompanied by a schema which defines the names of the tuples' fields. Tuples can contain primitive values such as integers, longs, shorts, bytes, strings, doubles, floats, booleans, and byte arrays. Additionally, custom serializers can be defined to interpret this data.
- The processing stages of a stream are defined in 'bolts' which can perform data manipulation, filtering, aggregations, joins, and so on. Bolts can also constitute more complex transforming structures that require multiple steps (thus, multiple bolts). The bolts can communicate with external applications such as databases and Kafka queues (Toshniwal et al., 2014).

In comparison to MapReduce and iterative algorithms introduced in the subsections above, Storm topologies, once created, run indefinitely until killed. Given this, the inefficient scattering of application's tasks among Cluster nodes has a lasting impact on performance. Storm's default scheduler implements a Round Robin strategy. For resource allocation purposes, Storm assumes that every worker is homogenous. This design results in frequent resource over-allocation and inefficient use of inter-system communications (Kulkarni et al., 2018). To try and solve this issue, more complex solutions are proposed such as D-Storm (Liu and Buyya, 2017). D-Storm's scheduling strategy is based on a metaheuristic algorithm Greedy, which also monitors the volume of the incoming workload and is resource-aware.

Typical examples of Storm's usage include:

- processing a stream of new data and updating databases in real time, for example in trading systems wherein data accuracy is crucial;
- continuously querying and forwarding the results to clients in real time, for example streaming trending topics on Twitter into browsers, and
- a parallelization of a computing-intensive query on the fly, i.e., a distributed Remote Procedure Call (RPC) wherein a large number of sets are probed (Marz, 2011).

Storm has gained widespread popularity and is used by companies such as Groupon, Yahoo!, Spotify, Verisign, Alibaba, Baidu, Yelp, and many more. A comprehensive list of users is available at the storm.apache.org website.

At the time of writing, Storm is being replaced at Twitter by newer distributed stream processing engine – Heron (Kulkarni et al., 2018) which continues the DAG model approach, but focuses on various architectural improvements such as reduced overhead, testability, and easier access to debug data.

## 5. GOOGLE'S BORG

To support its operations, Google utilizes a high number of data centers around the world, which at the time of writing number sixteen. Borg admits, schedules, starts, restarts and monitors the full range of applications run by Google. Borg users are Google developers and system administrators, and users submit their workload in the form of jobs. A job may consist of one or more tasks that all run the same program (Burns et al., 2016).

## 5.1. DESIGN CONCEPTS

The central module of the Borg architecture is BorgMaster, which maintains an in-memory copy of most of the state of the cell. This state is also saved in a distributed Paxos-based store (Lamport, 1998). While BorgMaster is logically a single process, it is replicated five times to improve fault-tolerance. The main design priority of Borg was resilience rather than performance. Google services are seen as very durable and reliable, the result of multi-tier architecture, where no component is a single point of failure exists. Current allocations of tasks are saved to Paxos-based storage, and the system can recover even if all five BorgMaster instances fail. Each cell in the Google Cluster in managed by a single BorgMaster controller. Each machine in a cell runs BorgLet, an agent process responsible for starting and stopping tasks and also restarting them should they fail. BorgLet manages local resources by adjusting local OS kernel settings and reporting the state of its node to the BorgMaster and other monitoring systems.

The Borg system offers extensive options to control and shape its workload, including priority bands for tasks (i.e., monitoring, production, batch, and best effort), resources quota and admission control. Higher priority tasks can pre-empt locally-running tasks to obtain the resources which are required. The exception is made for production tasks which cannot be pre-empted. Resource quotas are part of admission control and are expressed as a resource vector at a given priority, for some time (usually months). Jobs with insufficient quotas are rejected immediately upon submission. Production jobs are limited to actual resources available to BorgMaster in a given cell. The Borg system also exposes a web-based interface called Sigma, which displays the state of all users' jobs, shows details of their execution history and, if the job has not been scheduled, also provides a 'why pending?' annotation where there is guidance about how to modify the job's resource requests to better fit the cell (Verma et al., 2015).

The dynamic nature of the Borg system means that tasks might be started, stopped and then rescheduled on an alternative node. Google engineers have created the concept of a static Borg Name Service (BNS) which is used to identify a task run within a cell and to retrieve its endpoint address. The BNS address is predominantly used by load balancers to transparently redirect RPC calls to the endpoint of a given task. Meanwhile, the Borg's resource reclamation mechanisms help to reclaim under-utilized resources from cell nodes for non-production tasks. Although in theory users may request high resource quotas for their tasks, in practice they are rarely fully utilized continuously. Instead, they have peak times of the day or are used in this way when coping with a denial-of-service attack. BorgMaster has routines that estimate resource usage levels for a task and reclaim the rest for low-priority jobs from the batch or the best effort bands (Verma et al., 2015).

## 5.2. JOBS SCHEDULERS

Early versions of Borg had a simple, synchronous loop that accepted jobs requests and evaluated on which node to execute them. The current design of Borg deploys several schedulers working in parallel – the scheduler instances use a shared state of the available resources, but the resource offers are not locked during scheduling decisions (optimistic concurrency control). Where there is a conflicting situation where two or more schedulers allocate jobs to the same resources, all the jobs involved are returned to the jobs queue (Schwarzkopf et al., 2013).

When allocating a task, Borg's scheduler scores a set of available nodes and selects the most feasible machine for this task. Initially, Borg implemented a variation of the Enhanced Parallel Virtual Machine algorithm (E-PVM) (Amir et al., 2000) for calculating the task allocation score. Although this resulted in the fair distribution of tasks across nodes, it also resulted in increased fragmentation and later difficulties when fitting large jobs which required the most of the node's resources or even the whole node itself. An opposite to the E-PVM approach is a best-fit strategy, which, in turn, packs tasks very tightly. The best-fit approach may result in the excessive pre-empting of other tasks running on the same node, especially when the user miscalculates the resources required, or when the application has frequent load spikes. The current model used by Borg's scheduler is a hybrid approach that tries to reduce resource usage gaps (Verma et al., 2015).

Borg also takes advantage of resources pre-allocation using 'allocs' (short for allocation). Allocs can be used to pre-allocate resources for future tasks to retain resources between restarting a task or to gather class-equivalent or related tasks, such as web applications and associated log-saver tasks, onto the same machine. If an alloc is moved to another machine, its tasks are also rescheduled.

One point to note is that, similar to MetaCentrum users (Klusáček and Rudová, 2010), Google's users tend to overestimate the memory resources needed to complete their jobs, to prevent jobs being killed due to exceeding the allocated memory. In over 90% of cases, users overestimate how many resources are required, which in certain cases can waste up to 98% of the requested resource (Moreno et al., 2013; Ray et al., 2017).

## 5.3. OPTIMISATIONS

Over the years, Borg design has acquired several optimizations, namely:

- Score caching – checking the node's feasibility and scoring it is a computation-expensive process. Therefore, scores for nodes are cached and small differences in the required resources are ignored;
- Equivalence classes – submitted jobs often consist of several tasks which use the same binary and which have identical requirements. Borg's scheduler considers such a group of tasks to be in the same equivalence class. It evaluates only one task per equivalence class against a set of nodes, and later reuses this score for each task from this group;
- Relaxed randomization – instead of evaluating a task against all available nodes, Borg examines machines in random order until it finds enough feasible nodes. It then selects the highest scoring node in this set.

While the Borg architecture remains heavily centralized, this approach does seem to be successful. Although this eliminates head-of-line job blocking problems and offers better scalability, it also generates additional overheads for solving resource collisions. Nevertheless, the benefits from better scalability often outweigh the incurred additional computation costs which arise when scalability targets are achieved (Schwarzkopf et al., 2013).

## 6. SUMMARY AND CONCLUSIONS

This paper has presented a taxonomy of available schedulers, ranging from early implementations to modern versions. Aside from optimizing throughput, different class schedulers have evolved to solve different problems. For example, while OS schedulers maximize responsiveness, Cluster schedulers focus on scalability, provide support a wide range of unique (often legacy) applications, and maintain fairness. Big Data schedulers are specialized to solve issues accompanying operations on large datasets, and their scheduling mechanisms are often extensively intertwined with programming language features.

Table 1 presents a comparison of the presented schedulers with their main features and deployed scheduling algorithms:

| Scheduler class | Requirements known pre-execution | Fault-tolerance mechanisms | Configuration | Common algorithms | Scheduling decision overhead | Design focus (aside throughput) |
|---|---|---|---|---|---|---|
| OS Schedulers | No | No | Simple (compile-time and runtime parameters) | CS, CQ, MLFQ, O(n), O(1), Staircase, WFQ | very low – low | - single machine<br>- NUMA awareness<br>- responsiveness<br>- simple configuration |
| Cluster Schedulers | Yes[1] | Yes | Complex (configuration files and GUI) | FCFS (backfilling and gang-scheduling), SJF, Best-Fit, Scoring Functions | low - high | - distributed nodes<br>- fairness<br>- complex sharing policy<br>- power consumption<br>- fault-tolerance |
| Big Data Schedulers | Yes[2] | Yes | Complex (configuration files and GUI) | Best-Fit, FCFS (locality and gang-scheduling), Greedy, Fair Scheduler, Round Robin | low - medium | - specialized frameworks<br>- parallelism<br>- distributed data storage<br>- massive data |

1. Cluster users are notorious in overestimating resources needed for the completion of their tasks, which results in cluster system job schedulers often over-allocating resources (Klusáček and Rudová, 2010; Moreno et al., 2013).
2. MapReduce jobs tend to have consistent resource requirements, i.e., in majority of cases, every 'map' task processes roughly the same amount of data (input data block size is constant), while 'reduce' task requirements shall be directly correlated to the size of returned data.

Table 1: Schedulers comparison

OS schedulers have evolved in such a way that their focus is on maximizing responsiveness while still providing good performance. Interactive processes which sleep more often should be allocated time-slices more frequently, while background processes should be allocated longer, but less frequent execution times. CPU switches between processes extremely rapidly which is why modern OS scheduling algorithms were designed with very low overhead (Wong et al., 2008; Pinel et al., 2011). Most end-users for this class of schedulers are non-technical. As such, those schedulers usually have a minimum set of configuration parameters (Groves et al., 2009).

OS scheduling was previously deemed to be a solved problem (Torvalds, 2001), but the introduction and popularization of multi-core processors by Intel (Intel Core™2 Duo) and AMD (AMD Phenom™ II) in the early 2000s enabled applications to execute in parallel. This meant that scheduling algorithms needed to be re-implemented to be efficient once more. Modern OS schedulers also consider NUMA properties when deciding which CPU core the task will be allocated to. Furthermore, the most recent research explores the potential application of dynamic voltage and frequency scaling technology in scheduling to minimize power consumption by CPU cores (Sarood et al., 2012; Padoin et al., 2014).

Given that it is hard to build a good universal solution which caters to the complexities of modern hardware, it is reasonable to develop the modular scheduler architecture suggested in Lozi et al. (2016).

Cluster schedulers have a difficult mission in ensuring 'fairness'. In this context, namely a very dynamic environment consisting of variety of applications, fairness means sharing cluster resources proportionally while simultaneously ensuring a stable throughput. Cluster systems tend to allow administrators to implement complex resource sharing policies with multiple input parameters (Adaptive Computing, 2002). Cluster systems implement extensive fault-tolerance strategies and sometimes also focus on minimizing power consumption (Lang and Patel, 2010). Surprisingly, it appears that the most popular scheduling approach is a simple FCFS strategy with variants of backfilling. However, due to the rapidly increasing cluster size, the current research focuses on parallelization, as seen with systems such as Google's Borg and Microsoft's Apollo.

Big Data systems are still rapidly developing. Nodes in Big Data systems fulfil the dual purposes of storing distributed file system parts and providing a parallel execution environment for system tasks. Big Data schedulers inherit their general design from the cluster system's jobs schedulers. However, they are usually much more specialized for the framework and are also intertwined with the programming language features. Big Data schedulers are often focused on 'locality optimization' or running a given task on a node where input data is stored or in the closest proximity to it.

The design of modern scheduling strategies and algorithms is a challenging and evolving field of study. While early implementations often used simplistic approaches, such as a CS, modern solutions use complex scheduling schemas. Moreover, the literature frequently mentions the need for a modular scheduler architecture (Vavilapalli et al., 2013; Lozi et al., 2016) which could customize scheduling strategies to hardware configuration or applications.

## REFERENCES

"Apache Aurora." Aurora. Available from: http://aurora.apache.org/ Retrieved December 5, 2018. Version 0.19.0.

"Marathon: A container orchestration platform for Mesos and DC/OS." Mesosphere, Inc. January 10, 2018. Available from: https://mesosphere.github.io/marathon/ Retrieved February 7, 2018.

"Maui Administrator's Guide." Adaptive Computing Enterprises, Inc. May 16, 2002. Available from: http://docs.adaptivecomputing.com/maui/pdf/mauiadmin.pdf Retrieved November 5, 2014. Version 3.2.

"Nomad - Easily Deploy Applications at Any Scale", HashiCorp. Available from: https://www.nomadproject.io Retrieved March 19, 2018. Version 0.7.1.

"Top500 List - November 2017". TOP500 Project. November, 2017. Available from: https://www.top500.org/lists/2017/11/ Retrieved November 17, 2017.

"TORQUE Resource Manager. Administration Guide 5.1.2." Adaptive Computing Enterprises, Inc. November 2015. Available from: http://docs.adaptivecomputing.com/torque/5-1-2/torqueAdminGuide-5.1.2.pdf Retrieved November 15, 2016.

Amir, Yair, Baruch Awerbuch, Amnon Barak, R. Sean Borgstrom, and Arie Keren. "An opportunity cost approach for job assignment in a scalable computing cluster." IEEE Transactions on parallel and distributed Systems 11, no. 7 (2000): 760-768.

Arpaci-Dusseau, Remzi H., and Andrea C. Arpaci-Dusseau. "Operating systems: Three easy pieces." Arpaci-Dusseau Books, 2015.

Barroso, Luiz André, Jeffrey Dean, and Urs Hölzle. "Web search for a planet: The Google cluster architecture." Micro, IEEE 23, no. 2 (2003): 22-28.

Becchetti, L, Stefano Leonardi, Alberto Marchetti-Spaccamela, Guido Schäfer, and Tjark Vredeveld. (2006) "Average-case and smoothed competitive analysis of the multilevel feedback algorithm." Mathematics of Operations Research 31, no. 1: 85-108.

Blagodurov, Sergey, Sergey Zhuravlev, Alexandra Fedorova, and Ali Kamali. "A case for NUMA-aware contention management on multicore systems." In Proceedings of the 19th international conference on Parallel architectures and compilation techniques, pp. 557-558. ACM, 2010.

Bode, Brett, David M. Halstead, Ricky Kendall, Zhou Lei, and David Jackson. "The Portable Batch Scheduler and the Maui Scheduler on Linux Clusters." In Annual Linux Showcase & Conference. 2000.

Bonald, Thomas, Laurent Massoulié, Alexandre Proutiere, and Jorma Virtamo. "A queueing analysis of max-min fairness, proportional fairness and balanced fairness." Queueing systems 53, no. 1 (2006): 65-84.

Boutin, Eric, Jaliya Ekanayake, Wei Lin, Bing Shi, Jingren Zhou, Zhengping Qian, Ming Wu, and Lidong Zhou. "Apollo: Scalable and Coordinated Scheduling for Cloud-Scale Computing." In *OSDI*, vol. 14, pp. 285-300. 2014.

Bu, Yingyi, Bill Howe, Magdalena Balazinska, and Michael D. Ernst. "HaLoop: Efficient iterative data processing on large clusters." Proceedings of the VLDB Endowment 3, no. 1-2 (2010): 285-296.

Bulpin, James R. "Operating system support for simultaneous multithreaded processors." No. UCAM-CL-TR-619. University of Cambridge, Computer Laboratory, 2005.

Burns, Brendan, Brian Grant, David Oppenheimer, Eric Brewer, and John Wilkes. "Borg, Omega, and Kubernetes." Communications of the ACM 59, no. 5 (2016): 50-57.

Campbell, Matthew. "Distributed Scheduler Hell." DigitalOcean. SREcon17 Asia/Australia. May 24, 2017.

Corbató, Fernando J., Marjorie Merwin-Daggett, and Robert C. Daley. "An experimental time-sharing system." In Proceedings of the May 1-3, 1962, spring joint computer conference, pp. 335-344. ACM, 1962.

Corbet, Jonathan. "The staircase scheduler." LWN.net. June 2, 2004. Available from: https://lwn.net/Articles/87729/ Retrieved September 25, 2017.

Corbet, Jonathan. "The Rotating Staircase Deadline Scheduler." LWN.net. March 6, 2007. Available from: https://lwn.net/Articles/224865/ Retrieved September 25, 2017.

Corbett, James C., Jeffrey Dean, Michael Epstein, Andrew Fikes, Christopher Frost, Jeffrey John Furman, Sanjay Ghemawat et al. "Spanner: Google's globally distributed database." ACM Transactions on Computer Systems (TOCS) 31, no. 3 (2013): 8.

Dean, Jeffrey, and Sanjay Ghemawat. "MapReduce: a flexible data processing tool." Communications of the ACM 53, no. 1 (2010): 72-77.

Drepper, Ulrich. "What every programmer should know about memory." Red Hat, Inc. 11 (2007): 2007.

Etsion, Yoav, and Dan Tsafrir. "A short survey of commercial cluster batch schedulers." School of Computer Science and Engineering, The Hebrew University of Jerusalem 44221 (2005): 2005-13.

Foster, Ian, and Carl Kesselman. "Globus: A metacomputing infrastructure toolkit." The International Journal of Supercomputer Applications and High Performance Computing 11, no. 2 (1997): 115-128.

Foster, Ian, Carl Kesselman, and Steven Tuecke. "The anatomy of the grid: Enabling scalable virtual organizations." The International Journal of High Performance Computing Applications 15, no. 3 (2001): 200-222.

Gabriel, Edgar, Graham E. Fagg, George Bosilca, Thara Angskun, Jack J. Dongarra, Jeffrey M. Squyres, Vishal Sahay et al. "Open MPI: Goals, concept, and design of a next generation MPI implementation." In European Parallel Virtual Machine/Message Passing Interface Users' Group Meeting, pp. 97-104. Springer Berlin Heidelberg, 2004.

Gentzsch, Wolfgang. "Sun grid engine: Towards creating a compute power grid." In Cluster Computing and the Grid, 2001. Proceedings. First IEEE/ACM International Symposium on, pp. 35-36. IEEE, 2001.

Ghemawat, Sanjay, Howard Gobioff, and Shun-Tak Leung. "The Google file system." In ACM SIGOPS operating systems review, vol. 37, no. 5, pp. 29-43. ACM, 2003.

Gog, I. "Dron: An Integration Job Scheduler." Imperial College London (2012).

Grimshaw, Andrew S. "The Mentat run-time system: support for medium grain parallel computation." In Distributed Memory Computing Conference, 1990., Proceedings of the Fifth, vol. 2, pp. 1064-1073. IEEE, 1990.

Grimshaw, Andrew S., William A. Wulf, James C. French, Alfred C. Weaver, and Paul Reynolds Jr. "Legion: The next logical step toward a nationwide virtual computer." Technical Report CS-94-21, University of Virginia, 1994.

Groves, Taylor, Jeff Knockel, and Eric Schulte. "BFS vs. CFS - Scheduler Comparison." The University of New Mexico, 11 December 2009.

Hamscher, Volker, Uwe Schwiegelshohn, Achim Streit, and Ramin Yahyapour. "Evaluation of job-scheduling strategies for grid computing." Grid Computing—GRID 2000 (2000): 191-202.

Hart, Johnson M. "Win32 systems programming." Addison-Wesley Longman Publishing Co., Inc., 1997.

Helland, Pat, and Harris Ed "Cosmos: Big Data and Big Challenges." Stanford University, October 26, 2011.

Hindman, Benjamin, Andy Konwinski, Matei Zaharia, Ali Ghodsi, Anthony D. Joseph, Randy H. Katz, Scott Shenker, and Ion Stoica. "Mesos: A Platform for Fine-Grained Resource Sharing in the Data Center." In NSDI, vol. 11, no. 2011, pp. 22-22. 2011.

Isard, Michael, Mihai Budiu, Yuan Yu, Andrew Birrell, and Dennis Fetterly. "Dryad: distributed data-parallel programs from sequential building blocks." In ACM SIGOPS operating systems review, vol. 41, no. 3, pp. 59-72. ACM, 2007.

Isard, Michael, Vijayan Prabhakaran, Jon Currey, Udi Wieder, Kunal Talwar, and Andrew Goldberg. "Quincy: fair scheduling for distributed computing clusters." In Proceedings of the ACM SIGOPS 22nd symposium on Operating systems principles, pp. 261-276. ACM, 2009.

Jackson, David, Quinn Snell, and Mark Clement. "Core algorithms of the Maui scheduler." In Workshop on Job Scheduling Strategies for Parallel Processing, pp. 87-102. Springer, Berlin, Heidelberg, 2001.

Jones, M. Tim. "Inside the Linux 2.6 Completely Fair Scheduler - Providing fair access to CPUs since 2.6.23" In IBM DeveloperWorks. December 15, 2009.

Kannan, Subramanian, Mark Roberts, Peter Mayes, Dave Brelsford, and Joseph F. Skovira. "Workload management with LoadLeveler." IBM Redbooks 2, no. 2 (2001).

Kay, Judy, and Piers Lauder. "A fair share scheduler." Communications of the ACM 31, no. 1 (1988): 44-55.

Klusáček, Dalibor, and Hana Rudová. "The Use of Incremental Schedule-based Approach for Efficient Job Scheduling." In Sixth Doctoral Workshop on Mathematical and Engineering Methods in Computer Science, 2010.

Klusáček, Dalibor, Václav Chlumský, and Hana Rudová. "Optimizing user oriented job scheduling within TORQUE." In SuperComputing The 25th International Conference for High Performance Computing, Networking, Storage and Analysis (SC'13). 2013.

Kolivas, Con. "linux-4.8-ck2, MuQSS version 0.114." -ck hacking. October 21, 2016. Available from: https://ck-hack.blogspot.co.uk/2016/10/linux-48-ck2-muqss-version-0114.html Retrieved December 8, 2016.

Krauter, Klaus, Rajkumar Buyya, and Muthucumaru Maheswaran. "A taxonomy and survey of grid resource management systems for distributed computing." Software: Practice and Experience 32, no. 2 (2002): 135-164.

Kulkarni, Sanjeev, Nikunj Bhagat, Maosong Fu, Vikas Kedigehalli, Christopher Kellogg, Sailesh Mittal, Jignesh M. Patel, Karthik Ramasamy, and Siddarth Taneja. "Twitter Heron: Stream processing at scale." In Proceedings of the 2015 ACM SIGMOD International Conference on Management of Data, pp. 239-250. ACM, 2015.

Lamport, Leslie. "The part-time parliament." ACM Transactions on Computer Systems (TOCS) 16, no. 2 (1998): 133-169.

Lang, Willis, and Jignesh M. Patel. (2010) "Energy management for mapreduce clusters." Proceedings of the VLDB Endowment 3, no. 1-2: 129-139.

Lewis, Ian, and David Oppenheimer. "Advanced Scheduling in Kubernetes". Kubernetes.io. Google, Inc. March 31, 2017. Available https://kubernetes.io/blog/2017/03/advanced-scheduling-in-kubernetes Retrieved January 4, 2018.

Litzkow, Michael J., Miron Livny, and Matt W. Mutka. "Condor-a hunter of idle workstations." In Distributed Computing Systems, 1988., 8th International Conference on, pp. 104-111. IEEE, 1988.

Liu, Xunyun, and Rajkumar Buyya. "D-Storm: Dynamic Resource-Efficient Scheduling of Stream Processing Applications." In Parallel and Distributed Systems (ICPADS), 2017 IEEE 23rd International Conference on, pp. 485-492. IEEE, 2017.

Lozi, Jean-Pierre, Baptiste Lepers, Justin Funston, Fabien Gaud, Vivien Quéma, and Alexandra Fedorova. "The Linux scheduler: a decade of wasted cores." In Proceedings of the Eleventh European Conference on Computer Systems, p. 1. ACM, 2016.

Marz, Nathan. "A Storm is coming: more details and plans for release." Engineering Blog. Twitter, Inc. August 4, 2011. Available from: https://blog.twitter.com/engineering/en_us/a/2011/a-storm-is-coming-more-details-and-plans-for-release.html Retrieved July 16, 2018.

McCullough, John C., Yuvraj Agarwal, Jaideep Chandrashekar, Sathyanarayan Kuppuswamy, Alex C. Snoeren, and Rajesh K. Gupta. "Evaluating the effectiveness of model-based power characterization." In USENIX Annual Technical Conf, vol. 20. 2011.

Moreno, Ismael Solis, Peter Garraghan, Paul Townend, and Jie Xu. "An approach for characterizing workloads in google cloud to derive realistic resource utilization models." In Service Oriented System Engineering (SOSE), 2013 IEEE 7th International Symposium on, pp. 49-60. IEEE, 2013.

Murray, Derek G., Malte Schwarzkopf, Christopher Smowton, Steven Smith, Anil Madhavapeddy, and Steven Hand. "CIEL: a universal execution engine for distributed data-flow computing." In Proc. 8th ACM/USENIX Symposium on Networked Systems Design and Implementation, pp. 113-126. 2011.

Naik, Nitin. "Building a virtual system of systems using Docker Swarm in multiple clouds." In Systems Engineering (ISSE), 2016 IEEE International Symposium on, pp. 1-3. IEEE, 2016.

Pabla, Chandandeep Singh. "Completely fair scheduler." Linux Journal 2009, no. 184 (2009): 4.

Padoin, Edson L., Márcio Castro, Laércio L. Pilla, Philippe OA Navaux, and Jean-François Méhaut. "Saving energy by exploiting residual imbalances on iterative applications." In High Performance Computing (HiPC), 2014 21st International Conference on, pp. 1-10. IEEE, 2014.

Pascual, Jose, Javier Navaridas, and Jose Miguel-Alonso. "Effects of topology-aware allocation policies on scheduling performance." In Job Scheduling Strategies for Parallel Processing, pp. 138-156. Springer Berlin/Heidelberg, 2009.

Pinel, Frédéric, Johnatan E. Pecero, Pascal Bouvry, and Samee U. Khan. "A review on task performance prediction in multi-core based systems." In Computer and Information Technology (CIT), 2011 IEEE 11th International Conference on, pp. 615-620. IEEE, 2011.

Pinheiro, Eduardo, Ricardo Bianchini, Enrique V. Carrera, and Taliver Heath. "Load balancing and unbalancing for power and performance in cluster-based systems." In Workshop on compilers and operating systems for low power, vol. 180, pp. 182-195. 2001.

Pop, Florin, Ciprian Dobre, Gavril Godza, and Valentin Cristea. "A simulation model for grid scheduling analysis and optimization." In Parallel Computing in Electrical Engineering, 2006. PAR ELEC 2006. International Symposium on, pp. 133-138. IEEE, 2006.

Ray, Biplob R., Morshed Chowdhury, and Usman Atif. "Is High Performance Computing (HPC) Ready to Handle Big Data?" In International Conference on Future Network Systems and Security, pp. 97-112. Springer, Cham, 2017.

Rodriguez, Maria Alejandra, and Rajkumar Buyya. "A taxonomy and survey on scheduling algorithms for scientific workflows in IaaS cloud computing environments." Concurrency and Computation: Practice and Experience 29, no. 8 (2017).

Sarood, Osman, Phil Miller, Ehsan Totoni, and Laxmikant V. Kale. ""Cool" Load Balancing for High Performance Computing Data Centers." IEEE Transactions on Computers 61, no. 12 (2012): 1752-1764.

Schwarzkopf, Malte, Andy Konwinski, Michael Abd-El-Malek, and John Wilkes. "Omega: flexible, scalable schedulers for large compute clusters." In Proceedings of the 8th ACM European Conference on Computer Systems, pp. 351-364. ACM, 2013.

Shreedhar, Madhavapeddi, and George Varghese. "Efficient fair queueing using deficit round robin." In ACM SIGCOMM Computer Communication Review, vol. 25, no. 4, pp. 231-242. ACM, 1995.

Singh, Ajit. "New York Stock Exchange Oracle Exadata – Our Journey." Oracle, Inc. November 17, 2017. Available from: http://www.oracle.com/technetwork/database/availability/con8821-nyse-2773005.pdf Retrieved June 28, 2018.

Sliwko, Leszek. "A Scalable Service Allocation Negotiation For Cloud Computing." Journal of Theoretical and Applied Information Technology, Vol.96. No 20, pp. 6751-6782, 2018.

Smanchat, Sucha, and Kanchana Viriyapant. "Taxonomies of workflow scheduling problem and techniques in the cloud." Future Generation Computer Systems 52 (2015): 1-12.

Smarr, Larry, and Charles E. Catlett. "Metacomputing." Grid Computing: Making the Global Infrastructure a Reality (2003): 825-835.

Thain, Douglas, Todd Tannenbaum, and Miron Livny. "Distributed computing in practice: the Condor experience." Concurrency and computation: practice and experience 17, no. 2-4 (2005): 323-356.

Torvalds, Linus "Re: Just a second …" The Linux Kernel Mailing List. December 15, 2001. Available from http://tech-insider.org/linux/research/2001/1215.html Retrieved September 27, 2017.

Toshniwal, Ankit, Siddarth Taneja, Amit Shukla, Karthik Ramasamy, Jignesh M. Patel, Sanjeev Kulkarni, Jason Jackson et al. "Storm @Twitter." In Proceedings of the 2014 ACM SIGMOD international conference on Management of data, pp. 147-156. ACM, 2014.

Tyagi, Rinki, and Santosh Kumar Gupta. "A Survey on Scheduling Algorithms for Parallel and Distributed Systems." In Silicon Photonics & High Performance Computing, pp. 51-64. Springer, Singapore, 2018.

Vavilapalli, Vinod Kumar, Arun C. Murthy, Chris Douglas, Sharad Agarwal, Mahadev Konar, Robert Evans, Thomas Graves et al. "Apache hadoop yarn: Yet another resource negotiator." In Proceedings of the 4th annual Symposium on Cloud Computing, p. 5. ACM, 2013.

Verma, Abhishek, Luis Pedrosa, Madhukar Korupolu, David Oppenheimer, Eric Tune, and John Wilkes. "Large-scale cluster management at Google with Borg." In Proceedings of the Tenth European Conference on Computer Systems, p. 18. ACM, 2015.

White, Tom. "Hadoop: The definitive guide." O'Reilly Media, Inc. 2012.

Wong, C. S., I. K. T. Tan, R. D. Kumari, J. W. Lam, and W. Fun. "Fairness and interactive performance of O(1) and cfs linux kernel schedulers." In Information Technology, 2008. International Symposium on, vol. 4, pp. 1-8. IEEE, 2008.

Vohra, Deepak. "Scheduling pods on nodes. " In Kubernetes Management Design Patterns, pp. 199-236. Apress, Berkeley, CA. 2017.

Vagata, Pamela, and Kevin Wilfong. "Scaling the Facebook data warehouse to 300 PB." Facebook, Inc. April 10, 2014. Available from: https://code.fb.com/core-data/scaling-the-facebook-data-warehouse-to-300-pb/ Retrieved June 28, 2018.

Yoo, Andy B., Morris A. Jette, and Mark Grondona. "Slurm: Simple linux utility for resource management." In Workshop on Job Scheduling Strategies for Parallel Processing, pp. 44-60. Springer, Berlin, Heidelberg, 2003.

Yu, Jia, and Rajkumar Buyya. "A taxonomy of scientific workflow systems for grid computing." ACM Sigmod Record 34, no. 3 (2005): 44-49.

Zaharia, Matei, Dhruba Borthakur, J. Sen Sarma, Khaled Elmeleegy, Scott Shenker, and Ion Stoica. Job scheduling for multi-user mapreduce clusters. Vol. 47. Technical Report UCB/EECS-2009-55, EECS Department, University of California, Berkeley, 2009.

Zaharia, Matei, Mosharaf Chowdhury, Michael J. Franklin, Scott Shenker, and Ion Stoica. "Spark: Cluster computing with working sets." HotCloud 10, no. 10-10 (2010): 95.

Zaharia, Matei, Mosharaf Chowdhury, Tathagata Das, Ankur Dave, Justin Ma, Murphy McCauley, Michael J. Franklin, Scott Shenker, and Ion Stoica. "Resilient distributed datasets: A fault-tolerant abstraction for in-memory cluster computing." In Proceedings of the 9th USENIX conference on Networked Systems Design and Implementation, pp. 2-2. USENIX Association, 2012.

Zakarya, Muhammad, and Lee Gillam. "Energy efficient computing, clusters, grids and clouds: A taxonomy and survey." Sustainable Computing: Informatics and Systems 14 (2017): 13-33.

Zecevic, Petar, and Marko Bonaci. "Spark in Action." (2016).

Zhang, Zhuo, Chao Li, Yangyu Tao, Renyu Yang, Hong Tang, and Jie Xu. "Fuxi: a fault-tolerant resource management and job scheduling system at internet scale." Proceedings of the VLDB Endowment 7, no. 13 (2014): 1393-1404.